**Determining Ultra-low Absorption Coefficients of Organic Semiconductors from the Sub-bandgap Photovoltaic External Quantum Efficiency**

*Christina Kaiser, Stefan Zeiske, Paul Meredith* and Ardalan Armin**

Christina Kaiser, Stefan Zeiske, Prof. Paul Meredith, Dr. Ardalan Armin
Department of Physics, Swansea University, Singleton Park SA2 8PP, Wales, UK
E-mail: ardalan.armin@swansea.ac.uk, paul.meredith@swansea.ac.uk

Keywords: absorption coefficient, external quantum efficiency, sub-bandgap states, organic solar cells and organic photodetectors

Energy states below the bandgap of a semiconductor, such as trap states or charge transfer states in organic donor-acceptor blends, can contribute to light absorption. Due to their low number density or ultra-small absorption cross-section, the absorption coefficient of these states is challenging to measure using conventional transmission/reflection spectrophotometry. As an alternative, the external quantum efficiency (EQE) of photovoltaic devices is often used as a representative of the absorption coefficient, where the spectral line shape of the EQE is considered to follow the absorption coefficient of the active layer material. In this work, it is shown that the sub-bandgap EQE is subject to thickness dependent low finesse cavity interference effects within the device – making this assumption questionable. A better estimate for the absorption coefficient is obtained when EQE spectra corresponding to different active layer thicknesses are fitted simultaneously for one attenuation coefficient using an iterative transfer matrix method. The principle is demonstrated for two model acceptor-donor systems (PCE12-ITIC and PBTTT-PC$_{71}$BM) and accurate sub-gap absorption coefficients are determined. This approach has particular relevance for both understanding sub-gap states and their utilization in organic optoelectronic devices.

# 1. Introduction

Electro-optical modelling has been extensively used to design and optimise the performance of optoelectronic devices such as solar cells[1,2], photodetectors[3,4] and thin-film light-emitting diodes (LEDs)[5]. For example, an accurate electro-optical model can help to increase the photocurrent produced by a solar cell[6], tune the spectral response of a photodetector[7] and increase the outcoupling efficiency of thin-film LEDs[8]. To perform these optimisations, a full knowledge of the complex refractive indices of all active and buffer layers of the device structure is required. For photon energies $hv$ above the bandgap $E_g$, the complex refractive index is often determined through transmission and reflection spectrophotometry or spectroscopic ellipsometry. The same experimental techniques become very challenging when applied to the sub-bandgap optical constants, particularly because for $hv < E_g$ the absorptance is typically orders of magnitude weaker than for $hv > E_g$. Throughout this work, the absorptance correctly refers to the ratio of absorbed to incident light power, whereas absorption is the physical process of photo-excitation through light-matter interaction. Precise sub-bandgap optical constants are, however, crucial for modelling a variety of semiconductor devices operating at energies below the bandgap such as near-infrared photodetectors[9,10] or light emitting exciplex LEDs[11]. In addition, the spectral line shape of the absorption coefficient for $hv < E_g$ is indicative of a variety of phenomena such as disorder in amorphous semiconductors and charge transfer absorption in organic semiconductors. Disorder-induced and charge transfer state absorptions are related to loss mechanisms in organic solar cells, and have been studied extensively to improve solar cell efficiencies.[12–14] Given the considerable interest in the sub-bandgap absorption coefficients, sensitive methods of measuring it have been developed. This includes photothermal deflection (PDS) spectroscopy as a powerful technique for measuring absorption coefficients down to 0.001 cm$^{-1}$ in the best case, but more commonly down to 1 cm$^{-1}$ in condensed matter phases.[15,16],[17],[18] Similar to ellipsometry, PDS requires the preparation of single layer samples and a rather complicated data analysis subject to model fitting. Another technique is Fourier Transform Photocurrent Spectroscopy (FTPS) that is based on recording the photocurrent action spectrum using a Fourier Transform Infrared spectrometer. FTPS is often used to determine low energy trap states in inorganic semiconductors.[19,20] One should note that photocurrent-based methods may be more sensitive than PDS but can only probe the absorptance of the photocurrent-generating species. In amorphous silicon, it has been shown that FTPS systematically underestimates the optical absorptance for $hv < 1.4$ eV when compared with PDS.[21,22] Similar to FTPS, the external quantum efficiency (EQE) is another photocurrent-based method commonly used to characterize the photocurrent response of solar cells and photodetectors over the visible (VIS) and near-infrared (NIR) spectral range. The EQE is the product of absorptance ($A$) of the active layer as a function of wavelength and the internal quantum efficiency (IQE), such that

$$\text{EQE}(\lambda) = A(\lambda) \times \text{IQE}. \qquad (1)$$

As a consequence of equation (1), the spectral line shape of the EQE must be directly transferrable to $A$ assuming that the IQE does not feature any substantial energy dependence. The latter assumption has been proven in technologically relevant high efficiency systems such as efficient bulk heterojunction[23–26] and perovskite[27] solar cells. Nevertheless, examples for donor-acceptor systems with an excitation energy dependent IQE have been shown recently.[28,29] If the spectral lineshape of the EQE changes with applied bias, the IQE is not spectrally flat and $A$ can be more precisely inferred from the EQE measured at the largest possible bias to ensure photocurrent saturation.[28,30]

The active layer absorptance spectrum $A(\lambda)$ depends on the absorption coefficient $\alpha(\lambda)$, but also the wave-optics of the multi-layer stack device that is the solar cell or the photodetector. In the case of an optically thin layer with $\alpha t \ll 1$ (layer thickness $t$) and in the limit of negligible cavity effects (such as a semiconductor film deposited on a transparent substrate), it is often assumed that $A \approx \alpha t$ for $hv < E_g$.[26,31–33] Therefore the spectral lineshape of the EQE should follow α via *A*. However, the assumption of $A \approx \alpha t$ is invalid in the presence of electrodes as the device forms a low-finesse cavity with an associated thickness dependence. The well-known cavity effects are typically observed as pronounced interference fringes in the above-gap EQE (where α*t* ~ 1) but can have a similar effect on the sub-gap EQE. The prominence of these cavity effects depends primarily on the relative difference between the individual layer refractive indices. A higher mismatch of the refractive index between the organic / transparent conducting electrode interface causes substantial Fresnel reflection making the respective electrode partially reflective. This can be quite a subtle effect, but important in regions of low attenuation coefficient and/or high dispersion in the refractive index – both circumstances encountered below the bandgap again making $A \approx \alpha t$ a questionable approximation. A better estimate for *A* is obtained by modelling the optical field distribution within a device, which requires knowledge of the optical constants (that is the real and imaginary components of the refractive index) for all materials constituting the layers of the device. In the so-called transfer matrix method, the electric field is described as a position- and wavelength-dependent matrix considering the absorption and reflection of every layer and interface.[34] $A(\lambda)$ of the active layer is then calculated as the sum of the position (*x*)-dependent modulus squared of the electric field $E(\lambda, x)$ over the layer thickness *t* multiplied by the absorption coefficient α and the refractive index *n*.

$$A(\lambda) = \alpha(\lambda) n(\lambda) \sum_{x=0}^{t} |E(\lambda, x)|^2 \qquad (2)$$

It is a relatively straightforward task to model a thin film solar cell when all optical constants are known but challenging when the sub-bandgap optical constants are unknown. In this work, we determined absorption coefficients as low as $10^{-2}$ cm$^{-1}$ in the sub-gap spectral range of two exemplary acceptor-donor blends often used in high efficiency organic solar cells. The full molecular structures of these model systems, PBTTT-PC$_{71}$BM and PCE12-ITIC are provided in the Supporting Information (SI). Solar cells with active layer thicknesses between 60 – 375 nm were fabricated to study the effect of thickness on the shape of the EQE spectra in the sub-gap spectral range. The absorption coefficients were obtained by numerically fitting for multiple EQE spectra and including thickness-dependent optical field simulations via an iterative transfer matrix method. We validated the modelled coefficients by simulating EQE spectra of devices with different thicknesses where we find a good agreement between the simulated and experimental results. It becomes clear that the EQE lineshape is strongly dependent on the active layer thickness and hence assuming that $A \approx \alpha t$ can only be true for a small range of thicknesses. In the PCE12-ITIC system, the interference-induced thickness dependence of EQE in the sub-gap regime is more pronounced due to the large refractive index of this material system. These results indicate that one cannot directly relate the sub-gap EQE spectra to the sub-gap absorption coefficient without performing appropriate optical simulations. This is particularly important in systems with larger refractive index such as the recently introduced non-fullerene donor/acceptor blends.

## 2. Results

### 2.1 Optical Constants

PBTTT-PC$_{71}$BM (1:4) and PCE12-ITIC (1:1) were chosen as model material systems for investigations of sub-gap ultra-low attenuation coefficients. PBTTT-PC$_{71}$BM is known for its extended sub-gap EQE shoulder due to charge transfer state (CT) absorption[35], while PCE12-ITIC shows a narrow CT state-related-EQE contribution that is barely distinguishable from the above-gap EQE caused by singlet excitations[33]. First, we obtained the optical constants from single layer samples of spin-coated material with thicknesses ranging between 100 – 400 nm on both glass and silicon substrates. Ellipsometry is considered as an indirect method for obtaining the optical constant, as it measures polarization parameters tan Ψ(λ) and cos Δ(λ), which are then globally fitted for the film thickness $t$ and refractive index $n$ in the transparent region between 1200 – 1600 nm ($\kappa \approx 0$, i.e., the Cauchy region). Thereafter, $n$ and $\kappa$ are fitted stepwise over the entire spectrum with fixed $t$. **Figure 1** shows the optical constants obtained for PCE12-ITIC and PBTTT-PC$_{71}$BM. We find that the ellipsometric data could be reliably determined in good agreement between the experimental data and the mathematical model (mean-squared error < 15) as well as high reproducibility between samples in the VIS spectral range. The dashed lines in Figure 1 indicate the sensitivity limit for determining $\kappa$, which roughly corresponds to the bandgap energy beyond which this method is not applicable to $\kappa$ (as seen from the divergence of $\kappa$ for different $t$). At the same time, the relative change in $n$ with $t$ in the Cauchy regime compared to $\kappa$ is insignificant, as $n$ lacks spectral features and converges to a constant value around 2. In summary, we used $n$ as obtained from ellipsometry for transfer matrix simulations in the entire spectral range but dismissed $\kappa$ for wavelengths beyond the dashed lines illustrated in Figure 1. We further determined the optical constants of indium tin oxide (ITO), MoO$_3$ and ZnO experimentally, while relying on reported tabulated data for Ag.[36] As a next step, electrically inverted photovoltaic cells with the structure ITO (105 nm) / ZnO (37 nm) / active layer / MoO$_3$ (7 nm) / Ag (130 nm) were fabricated with different active layer thicknesses (PCE12-ITIC: 60, 132, 215, 375 nm and PBTTT-PC$_{71}$BM: 60, 81, 110, 140, 190 nm).

## 2.2 Numerical determination of κ

The EQE spectra are used in the procedure depicted in the process flowchart of **Figure 2** to obtain ultra-low attenuation coefficients. The full protocol was implemented in MATLAB and is provided in the SI. The numerical basis for this process is the Nelder-Mead simplex algorithm that minimizes a non-linear and constrained objective function.[37] Importantly, the presented method does not depend on the algorithm used for the minimization. Other derivative-free algorithms like the Brent's method or Powell's conjugate direction method can also be employed.[38] To numerically determine the attenuation coefficient κ, two steps were followed as shown in Figure 2a and 2c: (1) Estimation of the precise active layer thickness for the single device (defined as the 'pixel') and (2) calculation of the attenuation coefficient κ under consideration of the pre-optimized thickness. First, the absorption $A$ of each device is modelled by the transfer matrix method. The experimental input data required for the modeling is the measured active layer thickness $t_0$ and the optical constants in the VIS spectral range that is from 400 – 600 nm for PBTTT-PC$_{71}$BM and 400 – 700 nm for PCE12-ITIC. Thereafter, IQE$_{VIS}$ and the mean (IQE$_{mean}$) are obtained using equation (1). From the available ellipsometric constants, values were chosen that resulted in the minimum deviations of the IQE$_{VIS}$ from its mean. The IQE of one exemplary PCE12-ITIC device (375 nm active layer layer) is depicted in Figure 1b showing a deviation of the IQE of no more than 13 % from its mean. Such deviations can be caused by thickness variations over the device area which are common for solution-processed layers and could even be seen with the naked eye for the 210 nm PBTTT-PC$_{71}$BM device. We therefore pre-optimized the active layer thickness $t$ within ± 10 nm boundaries around the experimental thickness $t_0$. To do so, the absorptance $A'$ is calculated from the IQE$_{mean}$ as $A' = \text{EQE} \times \text{IQE}_{mean}^{-1}$. Figure 2b shows that the absolute difference between $A'$ and $A$ is largest where the IQE deviates most from its mean. To match $A'$ to $A$, $t$ is numerically varied to minimize the function $(A - A')^2/A$. In the case of $A = A'$, IQE$_{VIS}$ would be spectrally flat, which is impossible to achieve given the experimental errors and uncertainties related to optical constants and thicknesses of other device layers. Nevertheless, the IQE spectra remains approximately flat within these unavoidable errors. IQE spectra of the devices under investigation are shown in **Figure S1-S2**. The experimental and the numerically optimized thicknesses of all the investigated devices are shown in **Table S1**. Once the optimized thicknesses and IQE$_{mean}$ for all devices are obtained, the absorption spectra below the bandgap can be estimated by $A'' = \text{EQE} \times \text{IQE}_{mean}^{-1}$ using the experimental EQE over the full spectral range. Importantly, $A''$ is now a matrix containing the estimated absorptance of all devices. Figure 2c shows the process of numerically obtaining κ. The fixed input comprises the absorption matrix $A''$, the experimental attenuation coefficient at the last wavelength within the VIS spectral range $κ(λ_{m-1})$, the attenuation coefficient $n$ over the full spectral range, and the numerically obtained thicknesses $t$. The optimization of κ proceeds wavelength by wavelength starting from the $λ_m$, where $λ_{m-1}$ is the last wavelength within the VIS spectral range. At $λ_m$, the transfer matrix calculates $A'''$ using $κ(λ_{m-1})$ as the initial guess of the attenuation coefficient. To further match $A'''$ to $A''(λ_m)$, the κ is varied between 0 and $κ(λ_{m-1}) + κ(λ_{m-1})/10$ with the goal to minimize the function $(A''(λ_m) - A''')^2 / A''(λ_m)$. Once a pre-defined number of iterations has been exceeded, $κ(λ_m)$ is fixed and the algorithm proceeds to the next wavelength $λ_{m+1}$ choosing $κ(λ_m)$ as the initial estimate for fitting $κ(λ_{m+1})$.

### 2.3 EQE simulations

A computer code for this method is provided in the supporting information with exemplar data. The presented method is applicable to a single or more devices of a different active layer thickness. However, inaccuracies in the determined thickness and small deviations from the flat IQE assumption will have a large impact on the numerically obtained attenuation coefficient if only one device is considered. Therefore, it will be more accurate to numerically fit devices with different thicknesses. From the globally obtained attenuation coefficient, the active layer absorptance can be simulated and then scaled by $IQE_{mean}$ to obtain the EQE. It is important to consider that electronic effects, such as the charge collection imbalance of thick junctions, are not considered in the optical model. **Figure 3a** and **3b** illustrate the absorption coefficients, the experimental EQE ($EQE_{exp}$) and simulated EQE ($EQE_{sim}$) spectra of PCE12-ITIC and PBTTT-PC$_{71}$BM. The agreement between simulation and experiment for different active layer thicknesses was quantified by calculating the relative percentage difference. $EQE_{sim}$ is typically within 40 % deviation around the absolute $EQE_{exp}$ despite the simulation covering 7 orders of magnitude. Moreover, the difference in EQE$_{exp}$ with active layer thickness in the sub-bandgap region is mostly larger than the deviation of the simulation around EQE$_{exp}$. Consequently, EQE$_{sim}$ spectra follow the overall trend of the thickness dependent EQE$_{exp}$ spectra. To exemplify this, the maximum difference in EQE of 60 nm and 215 nm PCE12-ITIC devices in the sub-bandgap region is 20-fold, whereas the fit-error for one thickness is at most 2 times the experimental EQE, i.e. a maximum relative percentage error of 100 %. The strongest deviations of EQE$_{sim}$ from EQE$_{exp}$ (65 % and 75 %) occur in the spectral range of the bandgap (210 nm PBTTT-PC$_{71}$BM and 60 nm PCE12-ITIC) and at the sensitivity limit of the EQE measurement (e.g. 60 nm PCE12-ITIC and 60 nm PBTTT-PC$_{71}$BM above 1300 nm). From a Gaussian fit to the absorption coefficient of PBTTT-PC$_{71}$BM, a charge transfer energy ($E_{CT}$) of 1.16 eV and a reorganization energy λ of the donor-acceptor complex of 163 meV were determined as shown in **Figure S3**. $E_{CT}$ of PBTTT-PC$_{61}$BM was previously reported to be 1.15 eV in a thin solar cell, where the active layer absorptance $A$ in the device was approximated by $A \approx \alpha t$.[32] As both $E_{CT}$ values are in excellent agreement, we performed the Gaussian fit on EQE$_{exp}$ spectra for different active layer thicknesses under the assumption of $A \approx \alpha t$. The 190 nm PBTTT-PC$_{71}$BM, could not be reliably fitted, while other devices with active layer thicknesses of 60, 81, 110 and 140 nm resulted in $E_{CT}$ and λ values ranging from 1.15 – 1.18 eV and 143 – 186 meV respectively (see SI **Table S2**). Given the dependence of the fit parameters on the chosen fitting range, an uncertainty of ± 10 meV for $E_{CT}$ and ± 20 meV for λ is estimated. To conclude this phase of the analysis, the assumption of $A \approx \alpha t$ leads to an insignificant error in determining $E_{CT}$ and reorganisation energy for the material system PBTTT-PC$_{71}$BM for layer thicknesses below 190 nm. For PCE12-ITIC devices, the thickness dependent EQE features above 1000 nm were identified as resonances of the device acting as a low-finesse cavity. The resonance modes are highly dependent on the active layer thickness as expected. To eliminate the possible effect of higher order harmonics derived from the illumination system (see **Experimental**), the EQE of the 190 nm device was remeasured with long-pass filters placed in front of the sample. No significant changes in the EQE were observed as depicted in **Figure S5**. We should also note that the recorded EQE spectra are all above the electrical noise level of the apparatus as shown in **Figure S5**. In Figure 3a and 3b, the numerically obtained $\alpha$ is scaled to compare its spectral lineshape to the EQE spectra of devices with different active layer thicknesses $t$. The assumption $A \approx \alpha t$, according to which the spectral lineshape of $\alpha$ follows $A$ and therefore the EQE, is not applicable to any PCE12-ITIC device without greatly over- or underestimating $\alpha$. To find the thickness at which this assumption is valid, we simulated the active layer absorptance $A$ using the conventional transfer matrix method over a broad range of thicknesses (see **Figure S4**). For PCE12-ITIC, the best fit was obtained for thicknesses between 130 - 170 nm, however, the assumption clearly fails for other thicknesses. This is mainly caused by a strong dispersion of $n$ for PCE12-ITIC (see refractive indices in Figure 1). The strong dispersion arises from the high absorption coefficient and its sharp bandgap resulting in larger Cauchy factors. Consequently, interference effects are more pronounced in PCE12-ITIC than in PBTTT-PC$_{71}$BM devices, which presumably also applies to other non-fullerene containing systems with similarly large absorption coefficients. For PBTTT-PC$_{71}$BM, $A \approx \alpha t$ can give a better estimate for the numerical $\alpha$ over a wide range of thicknesses. At the same time, the change in thickness between 60 – 150 nm also affects the spectral lineshape of the simulated absorptance. From these two examples, it becomes clear that active layer thickness plays a significant role in the observed spectral lineshape of the EQE and care must be taken to not mistake material properties with device properties.

**2.4 Trap states**

Figure 3 further indicates a broad feature in the absorption coefficient spectrum of PCE12-ITIC between 1000 – 1400 nm with an ultra-low absorption coefficient of 0.1 – 0.01 cm$^{-1}$. According to the empirically established relation $E_{CT} \approx qV_{OC} + 0.6$ eV, the CT state energy of PCE12-

ITIC is expected to be around 1.5 eV. This has been experimentally shown[33,39] and can be observed in Figure 3 as a shoulder in the absorption coefficient spectrum at 800 nm. The corresponding Gaussian-distributed absorptance of CT states occurs at much higher energies than the observed tail states. Importantly, the ability to determine absorption coefficients between 0.1 – 0.01 cm$^{-1}$, allows one to observe ultra-weak optical transitions that are typically associated with defect absorption in this spectral range.[21,22,40] Moreover, the tail of the absorption coefficient observed here can be easily fitted by an exponential function. Similar to FTPS measurements, the presented method determines exclusively photocurrent-generating trap states. Therefore, it possibly underestimates the optical absorption coefficient in the spectral range of defect absorption. However, the thus obtained "effective absorption coefficient" is advantageous for modelling solely photocurrent-based devices like solar cells and photodetectors, and to perform detailed balance analysis to determine the open circuit voltage of the solar cells and thermodynamic limit of the detectivity of the photodetectors.

**2.5 Minimal EQE**

Using the presented method, the sensitivity to the absorption coefficient depends solely on the ability to measure small photocurrents in the weakly absorbing spectral region. Figure S5a shows the EQE of PCE12-ITIC cells measured up to 1700 nm. Depending on the device noise current and measurement integration time, the noise floor is reached at different wavelengths, but always above 1400 nm. In Figure S5b, the current response of a 105 nm thick PCE12-ITIC device in the dark is depicted as the noise-equivalent EQE of the measurement system utilized in this work. We believe it is therefore possible to obtain $\alpha$ values down to $10^{-3}$ cm$^{-1}$ if the device noise is decreased via (for example) increasing the shunt resistances (i.e., lowering the thermal noise). Importantly, this $\alpha$ is only valid for charge-generating states. Another limitation of this method concerns finding the accurate optical constants via spectroscopic ellipsometry needed to determine the IQE. Anisotropy or thickness dependent morphology as well as birefringent optical properties can be very difficult to model. Moreover, it has been reported that some material systems show an excitation energy dependent IQE.[23,28] In that case the sub-bandgap IQE cannot be inferred from the above bandgap IQE making the method presented here not applicable. However, if the experimental EQE spectra for several thicknesses can be simulated reliably, the numerical absorption coefficient is valid for the material system under investigation and hence the assumption of an energy independent IQE applies.

## 3. Conclusion

In summary, we have presented a numerical and experimental approach to determine ultra-small absorption coefficients of sub-gap states from external quantum efficiency spectra and applied it to organic semiconductors-based solar cells. Due to non-negligible cavity effects, the spectral shape of the sub-bandgap EQE is influenced by interference fringes that are greater when the refractive index is more dispersive as exemplified with the PCE12-ITIC system. The thickness-dependent EQE spectra of PCE12-ITIC and PBTTT-PC$_{71}$BM devices were accurately predicted when modelling the device using the as-obtained absorption coefficients. While the charge transfer energy of the PBTTT-PC$_{71}$BM was determined as previously reported, sub-bandgap states were observed for the PCE12-ITIC blend, which are most likely linked to photocurrent generating trap states. The method presented here refines how to obtain the absorption coefficient from EQE spectra while offering a novel way to study sub-bandgap absorption with relatively low experimental overhead.

## 4. Experimental Section

*Device preparation:* On a pre-cleaned glass substrate with a pre-structured layer of 105 nm ITO, a 35 nm thin layer of ZnO was deposited via a sol-gel spin coating at 4000 rpm from a 0.5 M solution of zinc acetate dihydrate in ethanolamine and 2-methoxy ethanol (volume ratio 3:97). The as-spun layer was annealed at 150 °C for 10 min in air. The ZnO hole-transporting layer was followed by the active layer comprising either PBTTT-PC$_{71}$BM or PCE12-ITIC. For

PBTTT-PC$_{71}$BM as the active layer, PBTTT (Sigma-Aldrich, M$_w$ > 50 000 g mol$^{-1}$, PDI < 3)) and [6,6]-Phenyl C$_{71}$-butyric acid methyl ester (PC$_{71}$BM) was mixed in a weight ratio of 1:4 and dissolved in chloroform:1,2-dichlorobenzene (6:4 volume ratio). The solution was stirred overnight, filtered through a 0.2 µm PTFE filter in the cold and spin-coated at 65 °C for 60 s. For film thicknesses between 100 – 200 nm, the concentration of the solution was chosen to be 32 mg mL$^{-1}$ and the spin-speed is varied between 1000 – 3000 rpm. For layer thicknesses below 100 nm, the same solution was diluted to 20 mg mL$^{-1}$ and the spin-speed varied between 1000 – 1500 rpm. For PCE12-ITIC as the active layer, the polymer PCE12 (PBDB-T) and the acceptor ITIC were mixed in a weight ratio of 1:1 and dissolved in chlorobenzene at 50 °C overnight to form a 20 mg mL$^{-1}$ solution. 0.5 vol% diiodooctane was added prior to spin-coating, while keeping the solution at 50 °C throughout the deposition process. Film thicknesses between 60 – 375 nm were obtained by varying the spin-speed between 400 – 2700 rpm and subsequent thermal annealing at 160 °C for 10 min. The active layer was followed by 7 nm of MoO$_3$ and 130 nm of silver thermally evaporated under high vacuum. All devices were defined by the geometrical overlap of the bottom and the top contact that equals 15 mm$^2$. To avoid exposure to ambient conditions, the organic part of the device was covered by a small glass substrate glued on top.

*Ellipsometry:* The optical constants of spin-coated films were determined experimentally by spectroscopic ellipsometry (instrument: J.A. Woollam M-2000; software: CompleteEASE 5.23). Samples comprised a single layer of material spin-coated from the same solution as utilized for device fabrication on either quartz glass or silicon substrates.

*Sensitive EQE measurements:* A Spectrophotometer (LAMBDA 950, PerkinElmer) with an integrated monochromator was used as a light source spanning 400 - 1800 nm. The output light from the spectrometer was chopped at 273 Hz and focused onto a photodiode. The short-circuit current was derived from the device was fed to a current preamplifier (Femto DHCPA-100) before being analyzed with a lock-in amplifier (SR860 Stanford Research Systems). To resolve low photocurrents, the preamplifier was set to high amplification, while the time-constant of the lock-in amplifier was chosen to be 1 s in the strongly absorbing and 30 s in the weakly absorbing spectral range. For the calibration process, a NIST-calibrated silicon photodiode (for wavelength between 400 nm and 1100 nm) and a germanium photodiode (for wavelength between 780 nm and 1800 nm) from Newport were used.

**Supporting Information**

**Acknowledgements**

This work was funded through the Welsh Government's Sêr Cymru II Program 'Sustainable Advanced Materials' (Welsh European Funding Office − European Regional Development Fund). C.K. and S.Z. are recipients of a UKRI EPSRC Doctoral Training Program studentship. P.M. is a Sêr Cymru II Research Chair and A.A. a Rising Star Fellow also funded through the Welsh Government's Sêr Cymru II 'Sustainable Advanced Materials' Program (European Regional Development Fund, Welsh European Funding Office and Swansea University Strategic Initiative).

Received: ((will be filled in by the editorial staff))
Revised: ((will be filled in by the editorial staff))
Published online: ((will be filled in by the editorial staff))

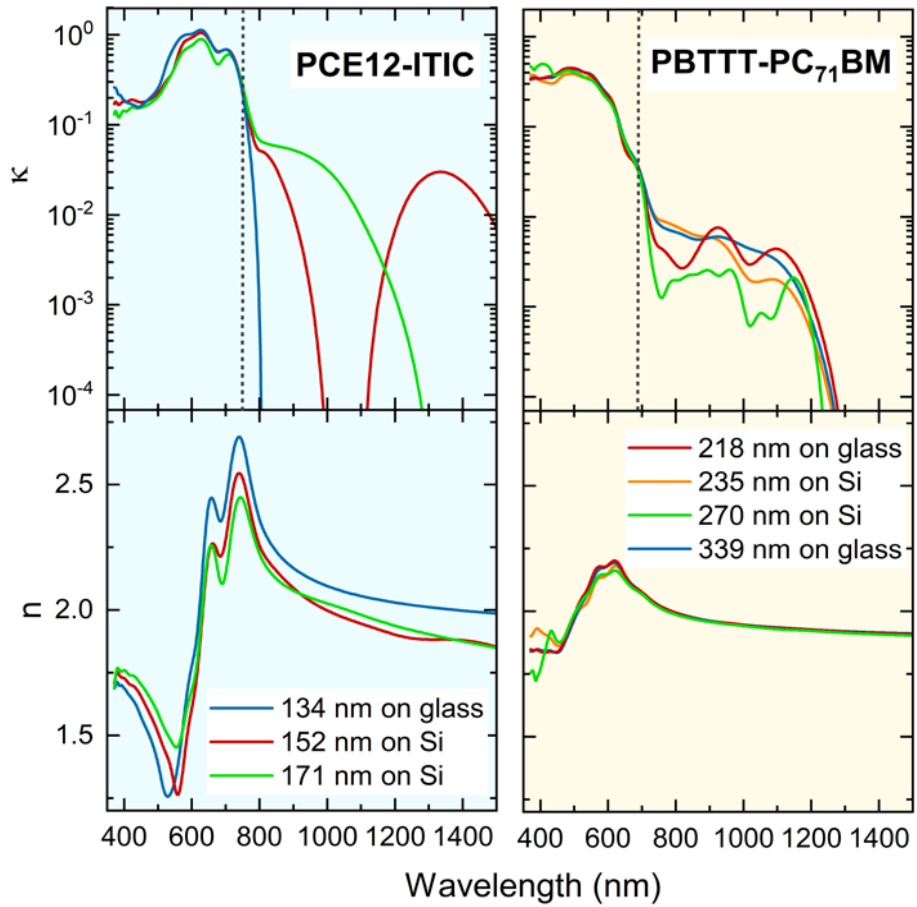

**Figure 1.** Spectroscopic ellipsometry obtained refractive indices *n* and attenuation coefficients κ of PBTTT-PC$_{71}$BM and PCE12-ITIC on silicon (Si) and glass with the respective layer thicknesses as shown. Dashed lines highlight the sensitivity limit in κ of the ellipsometric technique. PCE12-ITIC shows a significantly larger dispersion in *n* compared to PBTTT-PC$_{71}$BM.

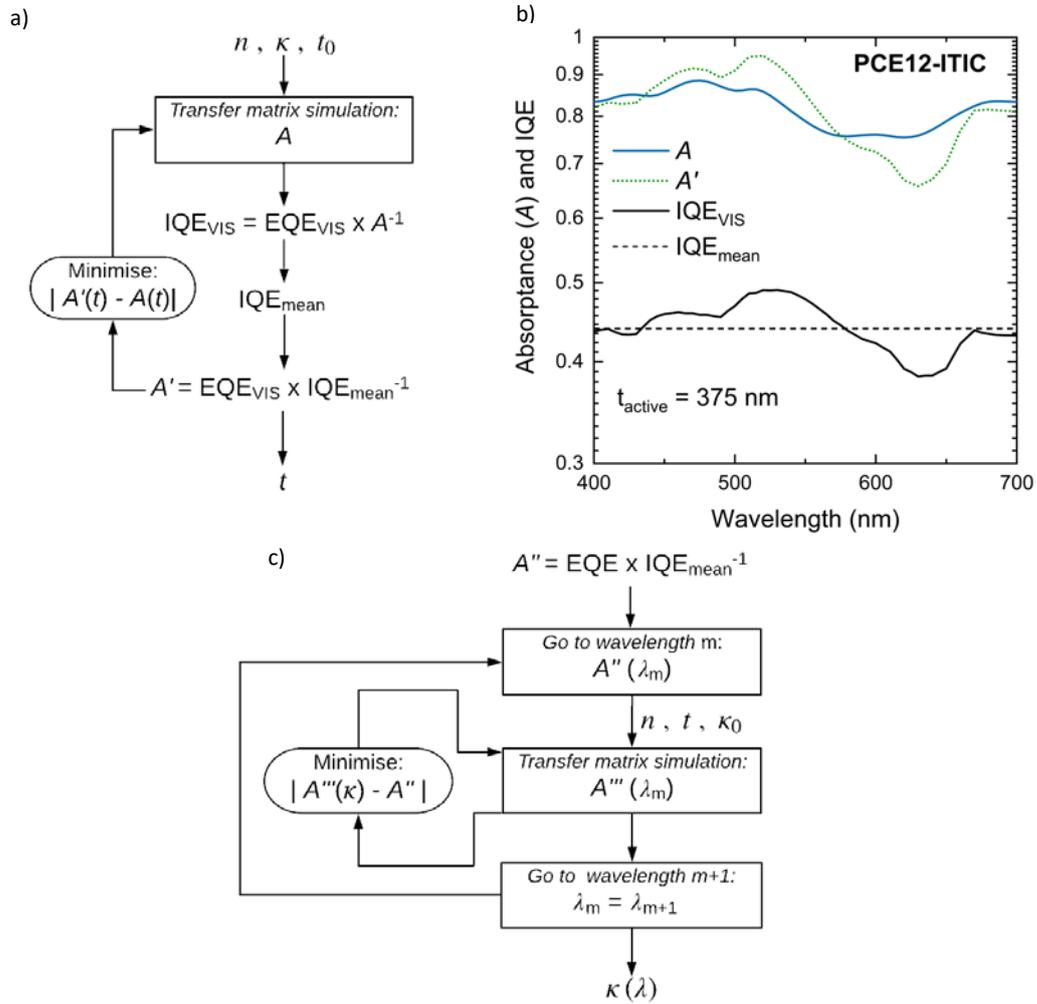

**Figure 2.** Process flow for obtaining ultra-low attenuation coefficients in the weakly absorbing spectral range using numerical optimization. a,b) The active layer thickness *t* is optimized such that the difference between $A$ and $A'$ in the visible spectral range is minimal resulting in a spectrally flat $IQE_{VIS}$. c) From the mean of the IQE and sensitively measured sub-bandgap EQE, an absorptance matrix $A''$ for several device thickness is calculated. Using the pre-optimized *t*, κ is numerically obtained by recursive simulation of the active layer absorptance $A'''(\lambda_m)$ for all device thicknesses at one wavelength with the goal to fit $A''$.

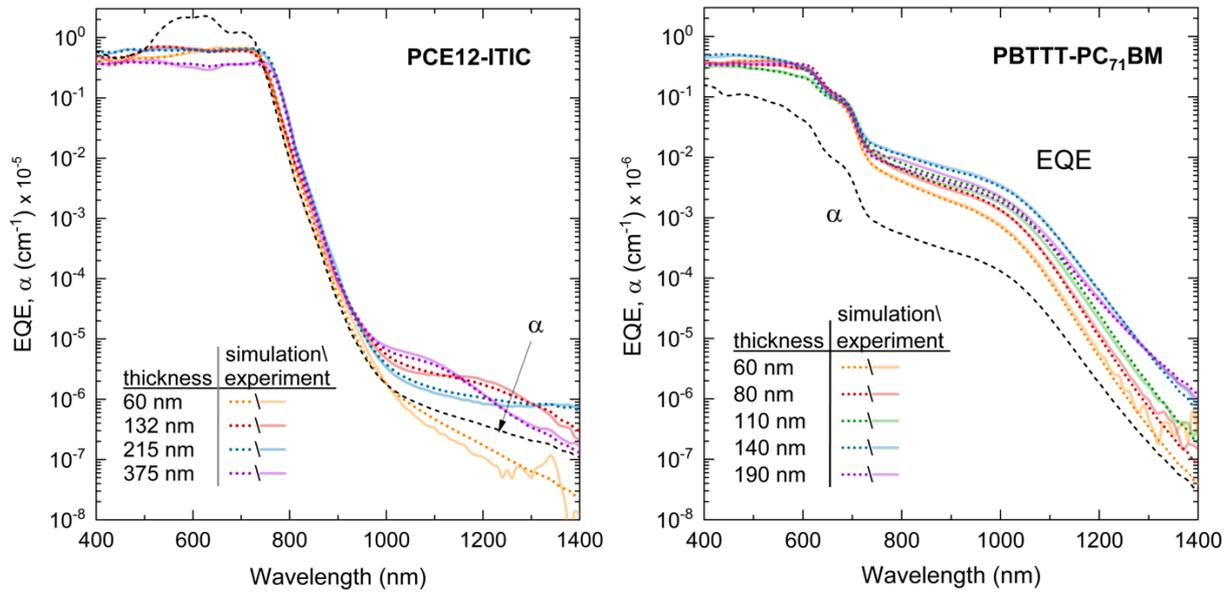

**Figure 3.** Absorption coefficients of PCE12-ITIC and PBTTT-PC$_{71}$BM (dashed black lines) comprising ellipsometric data for above-gap and numerical data for sub-bandgap absorption. Using $\alpha$ in a conventional transfer matrix simulation, the active layer absorptance was calculated and scaled by IQE$_{mean}$ to obtain the simulated EQE (dotted lines). The thickness-dependence of the sub-bandgap EQE is clearly reproduced in the simulation validating the numerical $\alpha$ as well as indicating that the commonly used assumption $At \approx \alpha$ is only true for a subset of thicknesses.

**ToC Text**: The absorption coefficient in the weakly absorptive region of the spectrum is of great importance for thin-film solar cells and photodiodes. This study presents a method for obtaining the sub-bandgap absorption coefficient from multiple EQE spectra.

*Christina Kaiser, Stefan Zeiske, Paul Meredith *, Ardalan Armin **

Determining Ultra-low Absorption Coefficients from the Sub-bandgap External Quantum Efficiency in Organic Solar Cells and Photodiodes

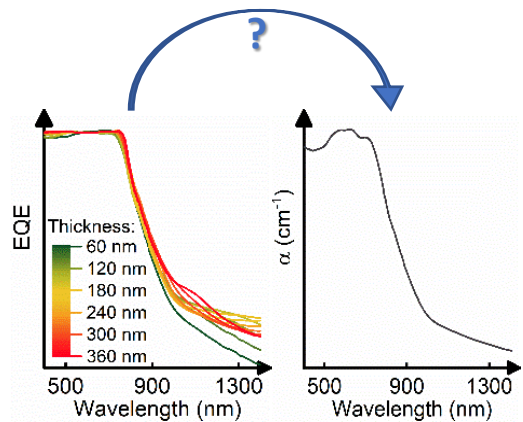

# Supporting Information

**Determining Ultra-low Absorption Coefficients from the Sub-bandgap External Quantum Efficiency in Organic Solar Cells and Photodiodes**

*Christina Kaiser, Stefan Zeiske, Paul Meredith \*, Ardalan Armin \**

Content:

Scheme S1: Molecular structures of PCE12, ITIC, PBTTT and $PC_{71}BM$

Figure S1: Active layer absorption and IQE of PCE12-ITIC devices in the VIS spectral

Figure S2: Active layer absorption and IQE of PBTTT-$PC_{71}BM$ devices in the VIS spectral

Table S1: Experimental and numerical active layer thicknesses

Figure S3: Gaussian fit to the absorption coefficient of PBTTT-$PC_{71}BM$

Table S2: Charge transfer energies and reorganization energies extracted from Gaussian fits to the experimental EQE spectra

Figure S4: Simulated sub-bandgap active layer absorption

Figure S5: EQE of PCE12-ITIC devices with long pass filters

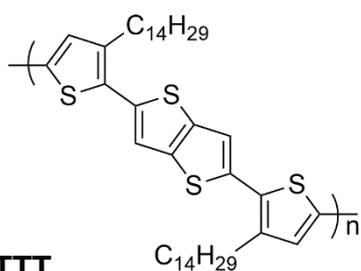

**PBTTT**

Poly[2,5-bis(3-tetradecylthiophen-2-yl)
thieno[3,2-*b*]thiophene]

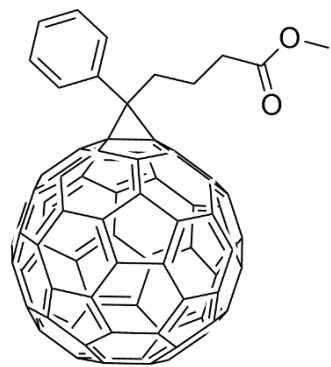

**PC71BM**

[6,6]-Phenyl C71 butyric acid
methyl ester

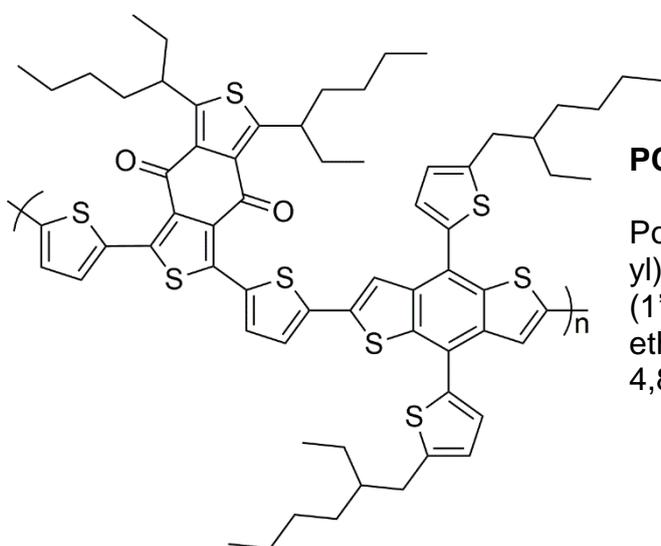

**PCE12**

Poly[(2,6-(4,8-bis(5-(2-ethylhexyl)thiophen-2-
yl)-benzo[1,2-b:4,5-b']dithiophene))-alt-(5,5-
(1',3'-di-2-thienyl-5',7'-bis(2-
ethylhexyl)benzo[1',2'-c:4',5'-c']dithiophene-
4,8-dione)]

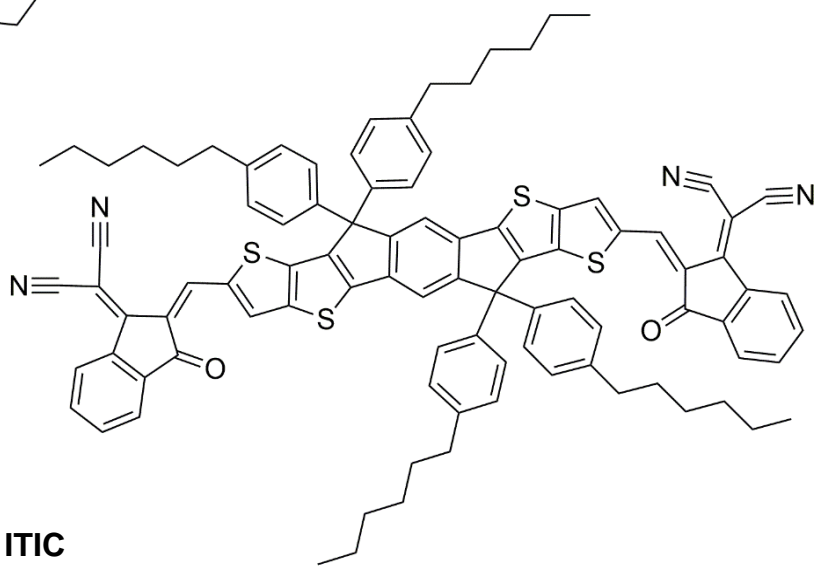

**ITIC**

3,9-bis(2-methylene-(3-(1,1-dicyanomethylene)-indanone))-
5,5,11,11-tetrakis(4-hexylphenyl)-dithieno[2,3-d:2',3'-d']-s-
indaceno[1,2-b:5,6-b']dithiophene

**Scheme S1** Molecular structures of the materials investigated in this study.

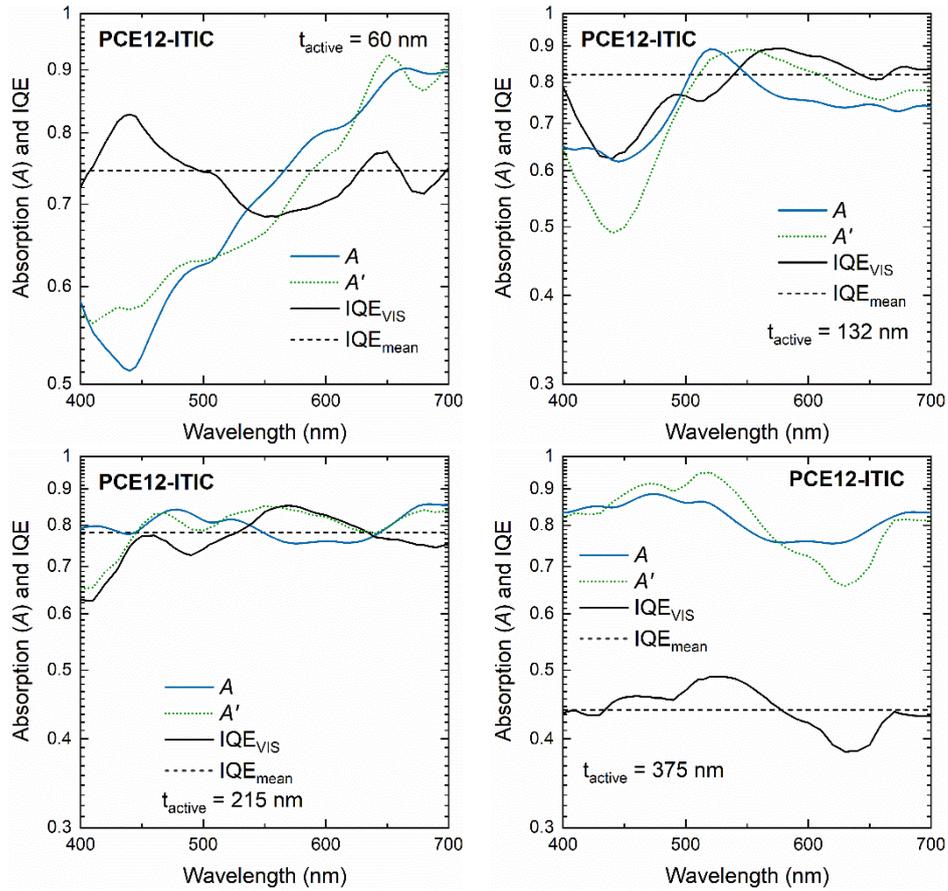

**Figure S1** Simulated active layer absorption, IQE and the mean of the IQE (IQE$_{mean}$) of devices with the structure ITO (109 nm) / ZnO (37 nm) / PCE12-ITIC (60 – 400 nm) / MoO$_3$ (7 nm) / Ag (130 nm). *A* corresponds to the absorption calculated by the transfer matrix using the experimentally obtained optical constants. *A'* corresponds to the absorption calculated as $A' = \text{EQE} \times \text{IQE}_{mean}^{-1}$.

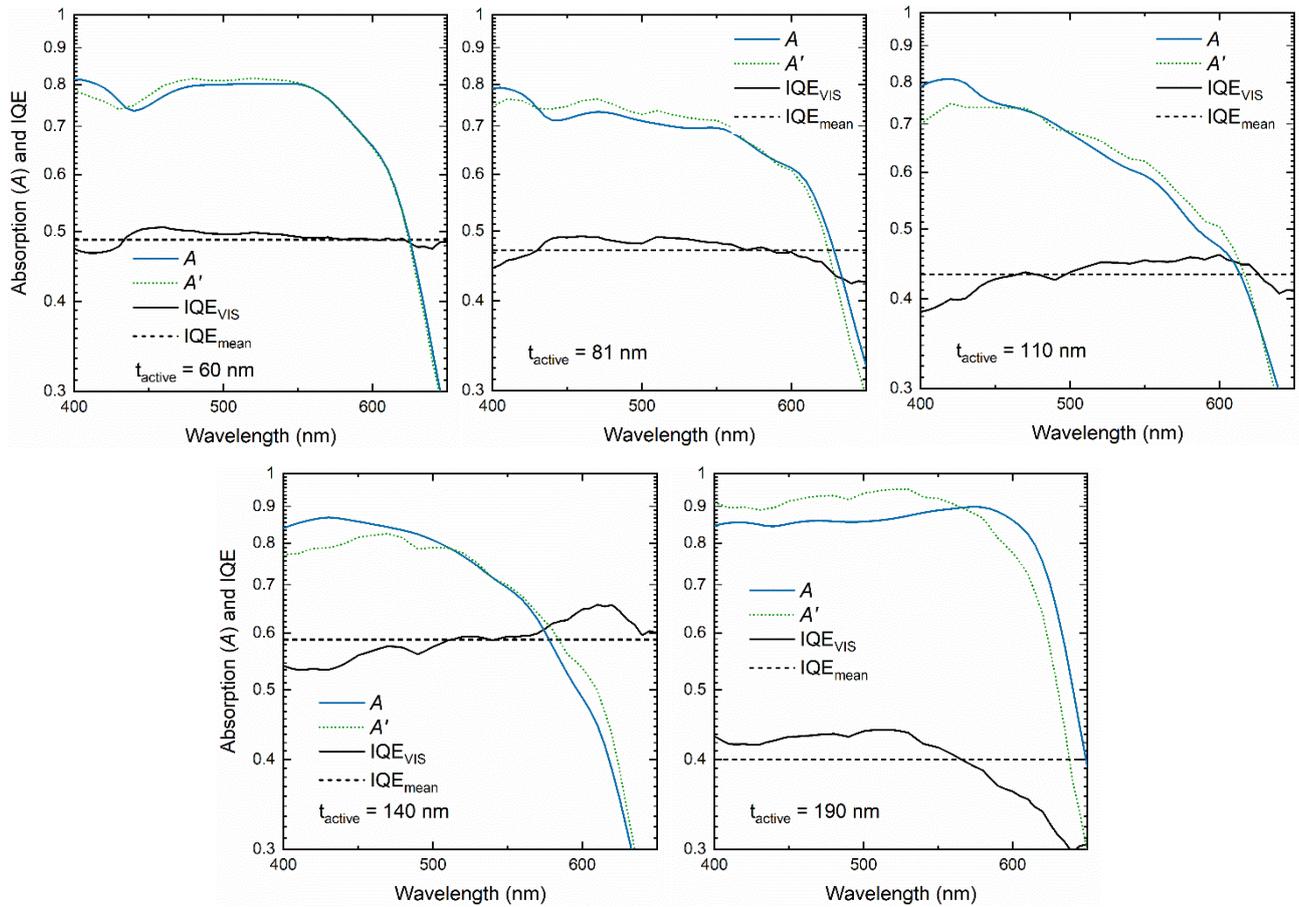

**Figure S2** Simulated active layer absorption, IQE and the mean of the IQE (IQE$_{mean}$) of a device with the structure ITO (109 nm) / ZnO (37 nm) / PBTTT-PC$_{71}$BM (60 – 190 nm) / MoO$_3$ (7 nm) / Ag (130 nm). *A* corresponds to the absorption calculated by the transfer matrix using the experimentally obtained optical constants. *A'* corresponds to the absorption calculated as $A' = EQE \times IQE_{mean}^{-1}$.

|  | Measured thickness in nm | Numerical thickness in nm |
|---|---|---|
| **PBTTT-PC$_{71}$BM** | 70 | 60 |
|  | 80 | 81 |
|  | 120 | 110 |
|  | 130 | 140 |
|  | 200 | 190 |
| **PCE12-ITIC** | 50 | 60 |
|  | 142 | 132 |
|  | 205 | 215 |
|  | 385 | 375 |

**Table S1** Active layer thickness of devices with either PBTTT-PC$_{71}$BM or PCE12-ITIC. The numerical thickness is obtained using a numerical fitting procedure that varies the thickness between ± 10 nm around the measured thickness. The fit aims to find a more precise thickness by reducing the deviations of the IQE from its mean.

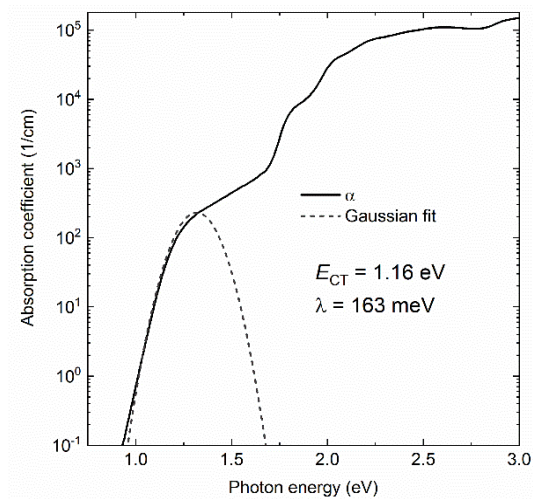

**Figure S3** Absorption coefficient of PBTTT-PC$_{71}$BM inferred from the ellipsometric extinction coefficient in the VIS spectral range and the numerical extinction coefficient in the sub-bandgap range. It is known that PBTTT-PC$_{71}$BM exhibits a strongly redshifted CT absorption leading to a sub-bandgap EQE shoulder. The low energy part of the absorption coefficient was fit by a Gaussian-like function to extract the charge transfer energy ($E_{CT}$) and the reorganization energy ($\lambda$). A charge transfer energy ($E_{CT}$) of 1.15 eV has already been reported in literature for PBTTT-PC$_{61}$BM.

|  | Active layer thickness in nm | $E_{CT}$ in eV | $\lambda$ in meV |
|---|---|---|---|
| **Absorption Coefficient** | -- | 1.155 | 162.7 |
| **EQE** | 60 | 1.175 | 152.9 |
|  | 81 | 1.181 | 134.4 |
|  | 110 | 1.150 | 182.5 |
|  | 140 | 1.158 | 150.8 |
|  | 190 | -- | -- |

**Table S2** Charge transfer energy ($E_{CT}$) and reorganization energy ($\lambda$) of PBTTT-PC$_{71}$BM obtained from a Gaussian fit to the low-energy part of the absorption coefficient or the experimental external quantum efficiency (EQE) spectra. The Gaussian fit to the EQE of the 190 nm thick device was not reliable and therefore excluded. $E_{CT}$ and $\lambda$ extracted as fit-parameter from the EQE spectra show negligible deviation from the values extracted from the absorption coefficient given an estimated fit-error of $\pm 10$ meV and $\pm 20$ meV for $E_{CT}$ and $\lambda$ respectively.

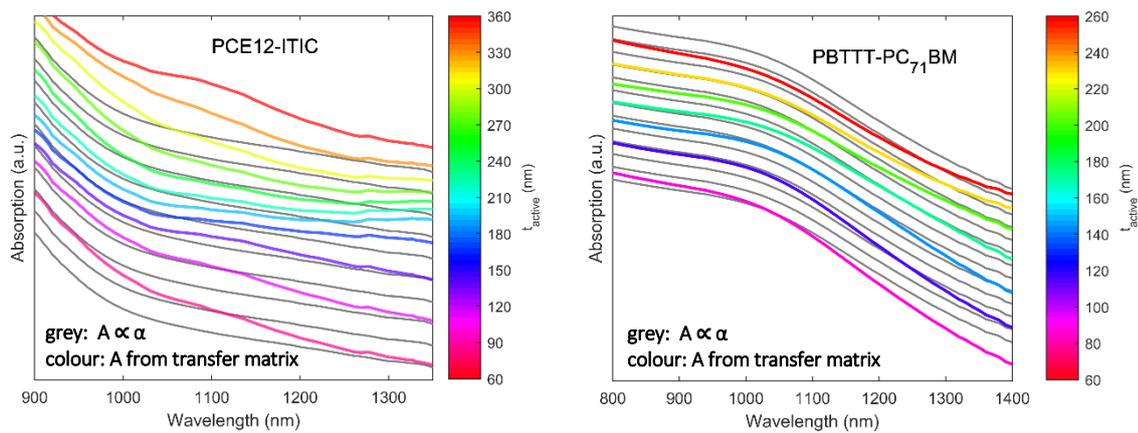

**Figure S4** Sub-bandgap spectral line-shape of the active layer absorption for different thicknesses in devices with PCE12-ITIC (left) and PBTTT-PC$_{71}$BM (right). Colored lines represent the simulated absorption obtained from the transfer matrix method and the numerical α. Grey lines represent the absorption obtained by assuming $A \approx \alpha t$ for random thicknesses.

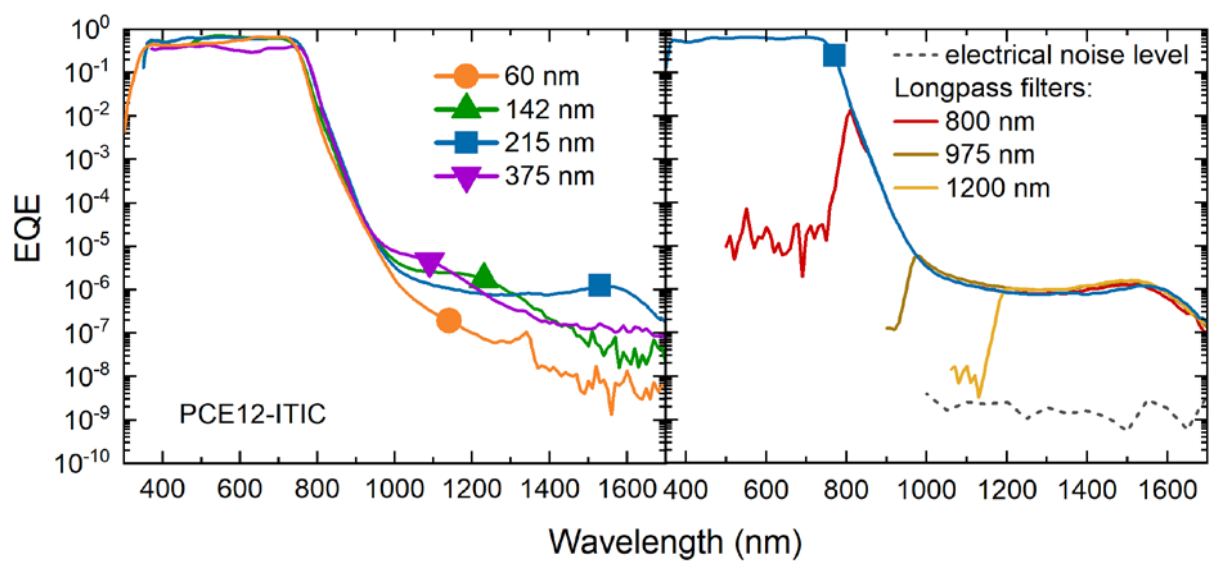

**Figure S5** Left: EQE spectra measured for PCE12-ITIC devices showing optical interference fringes as the result of the device acting as a low-finesse cavity for wavelengths above 1000 nm. Right: EQE of the PCE12-ITIC device with a 215 nm active layer. Different longpass filters were placed in front of sample to eliminate higher incident light order harmonics. The noise-equivalent EQE of a 105 nm thick PCE12-ITIC cell was recorded by fully blocking the light beam.